\documentclass[12pt,a4paper]{article}

\usepackage[T1]{fontenc}
\usepackage[utf8]{inputenc}
\usepackage{amsmath}
\usepackage{amssymb}
\usepackage{hyperref}
\usepackage[final]{graphicx}
\usepackage{bbm,latexsym,epsfig}

\newcommand{\be}{\begin{equation}}
\newcommand{\ee}{\end{equation}}
\newcommand{\ba}{\begin{eqnarray}}
\newcommand{\ea}{\end{eqnarray}}
\newcommand{\no}{\nonumber \\}

\begin{document}

\title{
\normalsize \hfill DO-TH/12-40 \\
\normalsize \hfill CFTP/12-018
\\[8mm]
\LARGE Flavour models for TM$_1$ lepton mixing}

\author{
Ivo de Medeiros Varzielas$^{(1)}$\thanks{E-mail: ivo.de@udo.edu}
\ and
Lu\'\i s Lavoura$^{(2)}$\thanks{E-mail: balio@cftp.ist.utl.pt}
\\*[8mm]
$^{(1)} \!$
\small Technical University of Dortmund, Faculty of Physics \\
\small Otto-Hahn-Stra\ss e 4, D-44221 Dortmund, Germany
\\*[4mm]
$^{(2)} \!$
\small Technical University of Lisbon, Instituto Superior T\'ecnico, CFTP, \\
\small 1049-001 Lisboa, Portugal
\\*[7mm]
}

\maketitle

\abstract{
We present a framework for lepton flavour models
such that the first column of the lepton mixing matrix is
$\left( 2, -1, -1 \right)^T \! \left/ \sqrt{6} \right.$. 
We show that the flavour symmetry group adequate for this purpose is $S_4$.
Our models are based on a vacuum alignment that can be obtained
in a supersymmetric framework.
}

\section{Introduction}

The recent experimental discovery that the lepton
mixing angle $\theta_{13}$ is non\-zero~\cite{minos,chooz,dayabay,reno}
has rendered outdated quite a few
previous phenomenological \textit{Ans\"atze}.
Notably,
the tri-bimaximal mixing (TBM) \textit{Ansatz}~\cite{HPS}
cannot stand in the face of the evidence for both a nonzero $\theta_{13}$
and a non-maximal atmospheric mixing angle $\theta_{23}$
(the latter evidence is still disputable~\cite{forero,fogli,schwetz}).
The stage is thus set
for searches for alternative models and \textit{Ans\"atze}.
One interesting possibility is the embedding
of a $\mu$--$\tau$ interchange in a generalised CP
symmetry~\cite{grimus,ferreira,nishi};\footnote{See also
refs.~\cite{Feruglio:2012cw,Holthausen:2012dk}.}
this allows for a nonzero $\theta_{13}$ but keeps $\theta_{23}$ maximal.
Another possibility consists in substituting the stringent TBM \textit{Ansatz}
by a relaxed version of it,
in which either only the first column or only the second column
of the lepton mixing matrix is assumed to take its TBM form;
these possibilities have been named TM$_1$ and TM$_2$,
respectively,
in ref.~\cite{albright}.
There are many other possibilities,
like for instance various models
featuring `texture' zeroes in the lepton mass matrices,
the \textit{Ansatz}\/ of lepton mixing `anarchy'~\cite{anarchy},
and models based on various flavour symmetry groups like $A_4$
(\textit{e.g.}\ refs.~\cite{Holthausen:2012wz,Varzielas:2012ai}),
$S_4$
(\textit{e.g.}\ ref.~\cite{Feruglio:2012cw}),
$\Delta(27)$
(\textit{e.g.}\ refs.~\cite{deMedeirosVarzielas:2006fc,ma}),
and so on (for a recent review,
see ref.~\cite{Altarelli:2012ss}).

The problem with many \textit{Ans\"atze}\/
is grounding them on well-defined field-theoretical models,
which might hope to render those \textit{Ans\"atze}\/
stable under renormalisation.
In particular,
this has already been achieved for TM$_2$~\cite{trimaximal}.
A model for TM$_1$ based on the strategy of
`sequential dominance'
has been presented in ref.~\cite{Antusch:2011ic}.
It is the purpose of this paper to suggest a different framework
for models featuring TM$_1$. 

The plan of the paper is as follows.
In section~\ref{sec:tm1} we define TM$_1$
and review its phenomenological merits
and predictions.\footnote{This has also been done recently
in ref.~\cite{Rodejohann:2012cf},
where a justification of the TM$_1$ \textit{Ansatz}\/
through a particular breaking of an $A_4$ flavour symmetry
has also been attempted.}
In section~\ref{sec:model} we present our framework for TM$_1$ models
by assuming a specific vacuum alignment in $S_4$-based
models.
In section~\ref{sec:align} we justify the vacuum alignment
used in the previous section in the context of supersymmetric versions
of our models.
Section~\ref{sec:conc} summarises our achievements.
In appendix~A we make a brief review of the group $S_4$,
its irreducible representations,
and the tensor products thereof.
Appendix~B considers the constraints on the neutrino mass spectrum
ensuing from some of our models.

\section{TM$_1$ \label{sec:tm1}}

TM$_1$ is defined to be the situation where
the first column of the lepton mixing matrix $U$ is
\begin{equation}
u_1 = \frac{1}{\sqrt{6}}
\left( \begin{array}{c}
2 \\ -1 \\ -1
\end{array} \right).
\label{eq:TM1def}
\end{equation}
In the standard parametrisation of $U$,
\begin{equation}
U = \left( \begin{array}{ccc}
c_{12} c_{13} &
s_{12} c_{13} &
s_{13} e^{- i \delta} \\
- s_{12} c_{23} - c_{12} s_{23} s_{13} e^{i \delta} &
c_{12} c_{23} - s_{12} s_{23} s_{13} e^{i \delta} &
s_{23} c_{13} \\
s_{12} s_{23} - c_{12} c_{23} s_{13} e^{i \delta} &
- c_{12} s_{23} - s_{12} c_{23} s_{13} e^{i \delta} &
c_{23} c_{13}
\end{array} \right) P,
\label{eq:stdpar}
\end{equation}
where $c_i = \cos{\vartheta_i}$ and $s_i = \sin{\vartheta_i}$
for $i = 12, 13, 23$
and $P$ is a $3 \times 3$ diagonal unitary matrix,
the diagonal elements of which are the `Majorana phases'.
Since in TM$_1$ $\left| U_{e1} \right|^2 = c_{12}^2 c_{13}^2 = 2/3$,
\begin{equation}
\label{s12}
s_{12}^2 = 1 - \frac{2}{3 c_{13}^2} = \frac{1 - 3 s_{13}^2}{3  - 3 s_{13}^2},
\quad \mbox{hence} \quad
c_{12}^2 = \frac{2}{3  - 3 s_{13}^2}.
\end{equation}
Moreover,
since $1/6 = \left| U_{\mu 1} \right|^2 = \left| U_{\tau 1} \right|^2$,
\begin{equation}
\label{uvyti}
\left( c_{23}^2 - s_{23}^2 \right) \left( s_{12}^2 - c_{12}^2 s_{13}^2 \right)
+ 4 c_{23} s_{23} c_{12} s_{12} s_{13} \cos{\delta} = 0.
\end{equation}
Inserting into eq.~(\ref{uvyti}) the values of $c_{12}$ and $s_{12}$
in eqs.~(\ref{s12}),
\begin{equation}
\left( c_{23}^2 - s_{23}^2 \right) \left( 1 - 5 s_{13}^2 \right)
+ 4 \sqrt{2 \left( 1 - 3 s_{13}^2 \right)} c_{23} s_{23} s_{13} \cos{\delta} = 0.
\end{equation}
Thus,
\begin{equation}
\label{delta}
\cos{\delta} = \frac
{\left( 1 - 5 s_{13}^2 \right) \left( 2 s_{23}^2 - 1 \right)}
{4 s_{13} s_{23} \sqrt
{2 \left( 1 - 3 s_{13}^2 \right) \left( 1 - s_{23}^2 \right)}}.
\end{equation}
Equation~(\ref{s12}) predicts $s_{12}^2$ as a function of $s_{13}^2$.
Equation~(\ref{delta}) predicts $\cos{\delta}$
as a function of $s_{13}^2$
and $s_{23}^2$.\footnote{See also ref.~\cite{Rodejohann:2012cf}.
A related analysis can be found in ref.~\cite{Ge:2011qn}.}

We use the phenomenological data of ref.~\cite{fogli}.\footnote{The papers
of refs.~\cite{forero,schwetz} provide other phenomenological fits
to the data.
The TM$_1$ \textit{Ansatz}\/ is just as good for those fits.
Notice,
however,
that $\cos{\delta}$ changes sign if $s_{23}^2$ is allowed to be above 0.5.}
According to that paper,
the best-fit values of $s_{13}^2$ and $s_{23}^2$
are approximately 0.024 and 0.390,
respectively.
Therefore,
eqs.~(\ref{s12}) and (\ref{delta}) may be approximated by
\begin{eqnarray}
\label{vjufit}
s_{12}^2 &\approx& 0.317 - 0.700 \left( s_{13}^2 - 0.024 \right),
\\
\label{ivurot}
\cos{\delta} &\approx& - 0.470 + 11.7 \left( s_{13}^2 - 0.024 \right)
+ 4.49 \left( s_{23}^2 - 0.390 \right),
\end{eqnarray}
respectively.
The prediction for $s_{12}^2$ in eq.~(\ref{vjufit})
agrees very well with experiment and is,
moreover,
almost independent of the precise value
of $s_{13}^2$.\footnote{It is remarkable
that even though,
experimentally,
the relative error in $s_{12}^2$
is approximately half the one in $s_{13}^2$,
the TM$_1$ \textit{Ansatz}\/ allows one to predict $s_{12}^2$
from $s_{13}^2$ and not the converse.}
Equation~(\ref{delta}) predicts $\cos{\delta}$ to be negative
as long as $\theta_{23}$ is in the first octant,
but the prediction for the exact value of $\cos{\delta}$
is much less precise than the one for $s_{12}^2$.
Using the $1\sigma$ intervals of ref.~\cite{fogli},
\ba
& & s_{13}^2 \in \left[ 0.0216,\, 0.0266 \right],
\ s_{23}^2 \in \left[ 0.365,\, 0.410 \right]
\no &\Rightarrow&
\cos{\delta} \in \left[ -0.622,\, -0.359 \right],
\ea
while,
using the $3 \sigma$ intervals,
\ba
& & s_{13}^2 \in \left[ 0.0169,\, 0.0313 \right],
\
s_{23}^2 \in \left[ 0.331,\, 0.637 \right]
\no &\Rightarrow&
\cos{\delta} \in \left[ -0.918,\, 0.505 \right].
\ea
\section{The models}
\label{sec:model}

\subsection{The general framework}

We now discuss a theoretical framework that leads to TM$_1$.
In our framework we assume that there are only three light neutrinos
and that they are Majorana particles.
The charged-lepton mixing matrix
is supposed to be diagonalised by the unitary matrix
\be
\label{Uomega}
U_\omega = \frac{1}{\sqrt{3}} \left( \begin{array}{ccc}
1 & 1 & 1 \\ 1 & \omega & \omega^2 \\ 1 & \omega^2 & \omega
\end{array} \right),
\ee
where $\omega = \exp{\left( i 2 \pi / 3 \right)}$.
Then,
the lepton mixing matrix is
\be
U = U_\omega U_\nu,
\label{Ulframe}
\ee
where $U_\nu$ is the unitary matrix that diagonalises
the effective light-neutrino Majorana mass matrix $M_\nu$:
\be
\label{dnu}
U_\nu^T M_\nu U_\nu = \mathrm{diag} \left( m_1, m_2, m_3 \right) \equiv D_\nu,
\ee
where the $m_j$
($j = 1, 2, 3$)
are non-negative real.
The symmetric matrix $M_\nu$ is supposed to have an eigenvector
$\left( 0, 1, 1 \right)^T$.
The most general symmetric matrix with that feature
may be written in the form
\be
\label{mnu}
M_\nu = \left( \begin{array}{ccc}
a+2b & f & -f \\ f & a-b & d \\ -f & d & a-b
\end{array} \right).
\ee
Then,
\be
\label{unu}
U_\nu = \left( \begin{array}{ccc}
0 & c e^{i \beta} & s e^{i \beta} \\ r & r s & - r c \\ r & - r s & r c
\end{array} \right) P,
\ee
where $r = 2^{-1/2}$,
$c = \cos{\sigma}$,
$s = \sin{\sigma}$,
and $P = \mathrm{diag} \left( e^{i \psi_1}, e^{i \psi_2}, e^{i \psi_3} \right)$.
Therefore,
from eqs.~(\ref{Ulframe}),
(\ref{Uomega}),
and~(\ref{unu}),
\be
\label{u}
U = \left( \begin{array}{ccccc}
\sqrt{2/3} & &
c e^{i \beta} \left/ \sqrt{3} \right. & &
s e^{i \beta} \left/ \sqrt{3} \right. \\
- \sqrt{1/6} & &
c e^{i \beta} \left/ \sqrt{3} \right. + i s \left/ \sqrt{2} \right. & &
s e^{i \beta} \left/ \sqrt{3} \right. - i c \left/ \sqrt{2} \right. \\
- \sqrt{1/6} & &
c e^{i \beta} \left/ \sqrt{3} \right. - i s \left/ \sqrt{2} \right. & &
s e^{i \beta} \left/ \sqrt{3} \right. + i c \left/ \sqrt{2} \right.
\end{array} \right) P,
\ee
which clearly satisfies the definition of TM$_1$ in eq.~(\ref{eq:TM1def}).
Comparing with the standard parametrisation in eq.~(\ref{eq:stdpar}),
one obtains
\ba
\label{s} s^2 &=& 3 s_{13}^2, \\
\label{beta} c s \sin{\beta} &=& \frac{\sqrt{3}}{2 \sqrt{2}}\,
c_{13}^2
\left( c_{23}^2 - s_{23}^2 \right).
\ea

At this stage it is useful to define the unitary matrices
\ba
O_1 &=& \left( \begin{array}{ccc}
0 & 1 & 0 \\ r & 0 & -r \\ r & 0 & r
\end{array} \right),
\\
O_2 &=& \left( \begin{array}{ccc}
1 & 0 & 0 \\ 0 & c e^{i \beta} & s e^{i \beta} \\ 0 & - s & c
\end{array} \right),
\ea
which are such that $U_\nu = O_1 O_2 P$.
Equation~(\ref{dnu}) may now be rewritten
\be
\label{main}
O_1^T M_\nu O_1 = O_2^\ast \
{\mathrm{diag}} \left( \mu_1, \mu_2, \mu_3 \right) \, O_2^\dagger,
\ee
where $\mu_j = m_j e^{- i 2 \psi_j}$.
Explicitly computing each matrix element on both sides of eq.~(\ref{main}),
one obtains
\begin{subequations}
\label{abd}
\ba
a - b + d &=& \mu_1, \\
a + 2 b &=& e^{- i 2 \beta} \left( c^2 \mu_2 + s^2 \mu_3 \right), \\
a - b - d &=& s^2 \mu_2 + c^2 \mu_3, \\
\sqrt{2} f &=& e^{- i \beta} c s \left( \mu_2 - \mu_3 \right).
\label{f}
\ea
\end{subequations}
One concludes from eq.~(\ref{f}) that $f \neq 0$ is mandatory,
lest either the matrix $O_2$ is trivial,
\textit{i.e.}\ $c s = 0$,
or the neutrinos $\nu_2$ and $\nu_3$ are degenerate;
both situations would contradict the phenomenology,
\textit{cf.}\ eq.~(\ref{s}).\footnote{In the most obvious situation,
$f=0$ would lead to $s_{13} = 0$ and,
thus,
to TBM.
Indeed,
our framework is simply an extension
of the framework of most models predicting TBM,
with the crucial difference that those models assume $f=0$
while we want $f$ to be nonzero.}

\subsection{Implementation of the framework with the group $S_4$}

We want the charged-lepton mass matrix to be diagonalised by $U_\omega$.
This materialises if that mass matrix is of the form
\be
\label{rightmass}
\left( \begin{array}{ccc}
\chi_1 & \chi_2 & \chi_3 \\
\chi_3 & \chi_1 & \chi_2 \\
\chi_2 & \chi_3 & \chi_1
\end{array} \right),
\ee
where $\chi_{1,2,3}$ are complex numbers which must be all different
lest the charged leptons
are massless.\footnote{Unfortunately, $\chi_{1,2,3}$ must be
severely finetuned in order for the charged-lepton masses to be
as hierarchical as observed experimentally.
For instance,
one possibility is that $\chi_1 = \chi_2 = \chi_3$ to very high
precision---this hypothesis is known in the literature as `flavour democracy'.
We note,
however,
that this finetuning is just as much of a problem in our framework as in
many other flavour models for lepton mixing
that do not rely on additional mechanisms
like the Froggatt--Nielsen paradigm~\cite{Froggatt:1978nt}.
Here we offer no solution to this conundrum.}
In order to obtain the matrix~(\ref{rightmass})
there must be a cyclic symmetry $D_{L1} \to D_{L2} \to D_{L3} \to D_{L1}$
among the three leptonic gauge-$SU(2)$ doublets $D_{L1,2,3}$.
Therefore,
the matrix
\be
G_3 = \left( \begin{array}{ccc}
0 & 1 & 0 \\ 0 & 0 & 1 \\ 1 & 0 & 0
\end{array} \right)
\ee
must represent one of the generators of the flavour symmetry group
in the representation to which the $D_{Lj}$ belong.

The neutrino mass matrix in eq.~(\ref{mnu}) is symmetric
under a transformation through the matrix
\be
F_3 = \left( \begin{array}{ccc}
1 & 0 & 0 \\ 0 & 0 & -1 \\ 0 & -1 & 0
\end{array} \right),
\ee
\textit{i.e.}\ $F_3 M_\nu F_3 = M_\nu$.
Therefore,
the matrix $F_3$ should represent another generator
of the flavour symmetry group
in the representation to which the $D_{Lj}$ belong.

The matrices $G_3$ and $F_3$ together generate the irreducible representation
$\mathbf{3}_1$ of the group $S_4$,
\textit{cf.}\ eq.~(\ref{G3F3}).
Therefore,
$S_4$ is the appropriate lepton flavour symmetry group
for a model predicting TM$_1$.
The group $S_4$ and its irreducible representations are reviewed in appendix~A.

We shall implement our models in a supersymmetric framework,
which is convenient in order to obtain the desired alignment
of vacuum expectation values (VEVs).
We allow for couplings of dimension higher than four,
adequately suppressed by as many powers as needed of a high-energy
(cutoff)
scale $\Lambda$.
We place both the $D_{Lj}$
and the charged-lepton gauge-$SU(2)$ singlets $\ell_{Rj}$
in representations $\mathbf{3}_1$ of the flavour symmetry group $S_4$.
The superpotential includes the gauge- and $S_4$-invariants
\begin{equation}
\frac{1}{\Lambda}\, H D_L \ell_R \left( y_1 T_1 + y_2 T_2 + y_S S \right),
\label{eq:PY}
\end{equation}
where $H$ is a gauge-$SU(2)$ doublet
which is invariant under the flavour symmetry $S_4$
and $y_{1,2,S}$ are coupling constants.
The superfields $T_1$,
$T_2$,
and $S$,
collectively known
as `familons' or `flavons',
are gauge singlets in distinct $S_4$ representations:
$T_1$ is a $\mathbf{3}_1$,
$T_2$ is a $\mathbf{3}_2$,
and $S$ is a $\mathbf{1}_1$,
\textit{i.e.}\ invariant under $S_4$.
In order to avoid undesirable terms,
we use an auxiliary Abelian symmetry,
which may be either a $U(1)$
or a $\mathbbm{Z}_N$ with sufficiently large $N$.
Let $c \left( f \right)$ denote the charge of the generic field $f$
under this $U(1)_\ell$ or $\mathbbm{Z}_{N\ell}$ symmetry.
It follows from eq.~(\ref{eq:PY}) that
$c \left( T_1 \right) = c \left( T_2 \right) = c \left( S \right)$
and that $c \left( H \right) + c \left( D_L \right)
+ c \left( \ell_R \right) = - c \left( S \right)$.
Also note that $c \left( T_1 \right) \neq 0$,
else a term with no flavon
(and no cutoff suppression)
should also be present in eq.~(\ref{eq:PY}).
Let $\left\langle 0 \left| T_1 \right| 0 \right\rangle =
\left( v_{1x}, v_{1y}, v_{1z} \right)$,
$\left\langle 0 \left| T_2 \right| 0 \right\rangle =
\left( v_{2x}, v_{2y}, v_{2z} \right)$,
and $\left\langle 0 \left| S \right| 0 \right\rangle = v_S$
denote the VEVs of the neutral-scalar components of these superfields,
then the charged-lepton mass matrix is
\be
\left( \begin{array}{ccccc}
y_S v_S & & y_1 v_{1z} + y_2 v_{2z} & & y_1 v_{1y} - y_2 v_{2y} \\
y_1 v_{1z} - y_2 v_{2z} & & y_S v_S & & y_1 v_{1x} + y_2 v_{2x} \\
y_1 v_{1y} + y_2 v_{2y} & & y_1 v_{1x} - y_2 v_{2x} & & y_S v_S
\end{array} \right).
\ee
This is of the form~(\ref{rightmass}) if
\be
\label{clalign}
v_{1x} = v_{1y} = v_{1z} \equiv v_1
\quad \mbox{and} \quad
v_{2x} = v_{2y} = v_{2z} \equiv v_2.
\ee
These are precisely the conditions for the breaking of the $S_4$ symmetry
in the charged-lepton sector to preserve the $\mathbbm{Z}_3$ symmetry
generated by $G_3$ alone,
\be
G_3
\left( \begin{array}{c} 1 \\ 1 \\ 1 \end{array} \right)
=
\left( \begin{array}{c} 1 \\ 1 \\ 1 \end{array} \right).
\ee
Note that the superfields $T_1$,
$T_2$,
and $S$ are all needed in order that the $\chi_j$ in eq.~(\ref{rightmass})
are all non-vanishing and different.

We give Majorana masses to the neutrinos through the superpotential terms
\be
\frac{1}{\Lambda}\, D_L \Delta D_L
\left( z_S \bar S + z_D \bar D + z_T \bar T \right),
\label{eq:Pnu}
\ee
where $\Delta$ is an $SU(2)$ triplet.
The neutrino mass matrix is generated via the VEV
of the neutral component of $\Delta$.
In eq.~(\ref{eq:Pnu}),
$\bar S$ is $S_4$-invariant,
$\bar D$ is a $\mathbf{2}$ of $S_4$,
$\bar T$ is a $\mathbf{3}_1$ of $S_4$,
and $z_{S,D,T}$ are their respective coupling constants.
Equation~(\ref{eq:Pnu}) is the most general $S_4$-invariant
since the symmetric part of the product of the two $\mathbf{3}_1$ of $S_4$
contains precisely an invariant,
a $\mathbf{2}$,
and a $\mathbf{3}_1$ of $S_4$,
\textit{cf.}\ eq.~(\ref{3times3}).
We introduce another auxiliary symmetry,
$U(1)_\nu$.\footnote{In principle it is possible
to combine the two auxiliary symmetries $U(1)_\ell$ and $U(1)_\nu$
into a single $U(1)$,
but that is beyond the scope of the present existence proof.
We note that by using different auxiliary symmetries
it is always possible to separate the charged-lepton and neutrino sectors,
since one of them involves the superpotential terms in eq.~(\ref{eq:PY})
while the other one involves the terms in eq.~(\ref{eq:Pnu}).}
Let $q \left( f \right)$ denote the charge under $U(1)_\nu$
of a generic superfield $f$.
Evidently all the flavons in eq.~(\ref{eq:Pnu}) must have the same
nonzero\footnote{If $q \left( \bar S \right)$ was zero,
terms without any flavon field would have to be added
to eq.~(\ref{eq:Pnu}).}
$U(1)_\nu$ charge,
\textit{i.e.}\ $q \left( \bar S \right)
= q \left( \bar D \right) = q \left( \bar T \right)$,
and that constitutes a serious constraint on the alignment mechanisms.
Let
$\left\langle 0 \left| \bar S \right| 0 \right\rangle = \bar v_S$,
$\left\langle 0 \left| \bar D \right| 0 \right\rangle =
\left( \bar v_{Dp}, \bar v_{Dq} \right)$,
and $\left\langle 0 \left| \bar T \right| 0 \right\rangle =
\left( \bar v_{Tx}, \bar v_{Ty}, \bar v_{Tz} \right)$,
then
\begin{subequations}
\ba
\left( M_\nu \right)_{11} &=&
z_S \bar v_S + z_D \left( \bar v_{Dp} + \bar v_{Dq} \right),
\\
\left( M_\nu \right)_{22} &=&
z_S \bar v_S + z_D \left( \omega \bar v_{Dp} + \omega^2 \bar v_{Dq} \right),
\\
\left( M_\nu \right)_{33} &=&
z_S \bar v_S + z_D \left( \omega^2 \bar v_{Dp} + \omega \bar v_{Dq} \right),
\ea
\end{subequations}
and
$\left( M_\nu \right)_{12} = z_T \bar v_{Tz}$,
$\left( M_\nu \right)_{13} = z_T \bar v_{Ty}$,
and $\left( M_\nu \right)_{23} = z_T \bar v_{Tx}$.
One sees that $M_\nu$ is of the desired form in eq.~(\ref{mnu})
if $\bar v_{Dp} = \bar v_{Dq} \equiv \bar v_D$
and $\bar v_{Ty} = - \bar v_{Tz}$;
this is precisely the alignment of VEVs that preserves
the $\mathbbm{Z}_2$ subgroup of $S_4$ generated,
in the representation $\mathbf{3}_1$,
by $F_3$:
\be
F_3
\left( \begin{array}{c} k \\ 1 \\ -1 \end{array} \right)
=
\left( \begin{array}{c} k \\ 1 \\ -1 \end{array} \right),
\quad
F_2
\left( \begin{array}{c} 1 \\ 1 \end{array} \right)
=
\left( \begin{array}{c} 1 \\ 1 \end{array} \right).
\ee
In the notation of the previous subsection,
we have $a = z_S \bar v_S$,
$b = z_D \bar v_D$,
$d = z_T \bar v_{Tx}$,
and $f = z_T \bar v_{Tz}$. 
Particular cases of interest occur when either $a$,
$b$,
or $d$ vanish;
this may be because some VEV vanishes or---in the cases
of $\bar S$ and $\bar D$---because one may altogether avoid
introducing those superfields
in a particular model.\footnote{We should not,
though,
omit both $\bar S$ and $\bar D$,
because that would lead to $a = b = 0$,
which cannot fit the experimental data.
See appendix~B for details.}
Those particular cases of interest are dealt with in appendix~B.

\section{Alignment}
\label{sec:align}

In this section we provide an existence proof of the alignments of VEVs
required in the previous section.
In this proof we assume a supersymmetric implementation
of the models,\footnote{In a full supersymmetric construction,
the relevant Yukawa terms require the use
of both two gauge-$SU(2)$ doublets and two gauge-$SU(2)$ triplets,
see refs.~\cite{Hambye:2000ui,Rossi:2002zb}.}
with an $R$-symmetry under which the Standard Model fermions
have $R$-charge $+1$.\footnote{We recall that all the
allowed superpotential terms must have $R$-charge $+2$.}
We rely mostly on $F$-terms
for selecting the alignment
directions,\footnote{A similar implementation of an analogous alignment
was discussed in refs.~\cite{Bazzocchi:2009pv, Dutta:2009bj, BhupalDev:2012nm}.
Here, however, we impose the
restriction that
the terms in the superpotential responsible for the alignment
must be \emph{renormalisable}.
Non-renormalisable terms are used only
in the Yukawa couplings by means of Froggatt--Nielsen \cite{Froggatt:1978nt}
messenger fields with the appropriate gauge representations.}
even though,
at least for some of the aligned directions,
we expect it to be possible to simplify the models
by employing also $D$-term alignment,
as done for instance in refs.~\cite{Varzielas:2008jm, Varzielas:2012ss}.

Before discussing specific alignments
it is useful to clarify some general points:
\begin{enumerate}
\item With the $F$-term alignment employed,
a set of specific directions is obtained for the minima.
As is often the case in this type of models,
physically distinct directions
are present;
we argue that Nature either happened to choose
the phenomenologically viable direction
out of a discrete number of degenerate choices,
or that some unspecified soft supersymmetry-breaking terms
lift the degeneracy, as argued in ref.~\cite{Bazzocchi:2009pv}.
We must therefore guarantee that the desired direction
is part of a \emph{discrete}\/ set of directions.
When discussing specific alignments
we shall highlight these degenerate directions.
\item The $F$-term alignment fixes directions for the VEVs
but does not fix their absolute value.
As in refs.~\cite{Bazzocchi:2009pv, Varzielas:2012ss},
we assume that the magnitude of the VEVs is nonzero
and that it is stabilised at some finite value;
this can be achieved through soft supersymmetry-breaking terms
including squared masses that become negative,
as argued in ref.~\cite{Varzielas:2012ss}.
\end{enumerate}

\subsection{Charged-lepton sector}

In this subsection we want to explain the alignment of eqs.~(\ref{clalign}).
We consider an interaction
\begin{equation}
P_T = T^0 \left( a_1 S T_1 + a_2 T_1 T_1 + a_3 T_2 T_2 + a_4 T_1 T_2 \right),
\label{eq:T1T2-}
\end{equation}
where $T^0$ is a `driving field' or `alignment field'
(as a matter of fact,
it is a set of three superfields)
which transforms as $\mathbf{3}_1$ under $S_4$,
has $R$-charge $+2$,
and has auxiliary charge $c \left( T^0 \right) = - 2 c \left( S \right)$.
The $a_{1,2,3,4}$ are coupling constants.
Taking the derivative of $P_T$ with respect
to the three components of $T^0$,
one obtains the $F$-terms
\begin{subequations}
\label{eq:FT0}
\begin{eqnarray}
\frac{\partial P_T}{\partial T^{0}_1} &=&
a_1 v_S v_{1x} + 2 a_2 v_{1y} v_{1z} + 2 a_3 v_{2y} v_{2z}
+ a_4 \left( v_{1y} v_{2z} - v_{1z} v_{2y} \right),
\nonumber \\ & & \\
\frac{\partial P_T}{\partial T^{0}_2} &=&
a_1 v_S v_{1y} + 2 a_2 v_{1z} v_{1x} + 2 a_3 v_{2z} v_{2x}
+ a_4 \left( v_{1z} v_{2x} - v_{1x} v_{2z} \right),
\nonumber \\ & & \\
\frac{\partial P_T}{\partial T^{0}_3} &=&
a_1 v_S v_{1z} + 2 a_2 v_{1x} v_{1y} + 2 a_3 v_{2x} v_{2y}
+ a_4 \left( v_{1x} v_{2y} - v_{1y} v_{2x} \right).
\nonumber \\ & &
\end{eqnarray}
\end{subequations}
In order to minimise the potential
we must set all three $F$-terms in eqs.~(\ref{eq:FT0}) to zero.
It is clear that there is a solution featuring eq.~(\ref{clalign}),
with
\begin{equation}
a_1 v_S v_1 + 2 a_2 v_1^2 + 2 a_3 v_2^2 = 0.
\label{clmagn}
\end{equation}
Here we have an instance of degenerate
directions for the VEVs as mentioned previously.
Namely,
it would also be possible to choose
\emph{both}\/ $\left( v_{1x}, v_{1y}, v_{1z} \right)$
and $\left( v_{2x}, v_{2y}, v_{2z} \right)$
to lie in the $\left(1, -1, 1 \right)$ direction,
provided\footnote{We note that
the direction $\left( 1, -1, 1 \right)$
is \emph{not}\/ related to the direction $\left( 1, 1, 1 \right)$
through an $S_4$ transformation,
since the matrix $\mathrm{diag} \left( 1, -1, 1 \right)$
does not belong to $S_4$.
Correspondingly,
eqs.~(\ref{clmagn}) and~(\ref{clmagn-}) are distinct.}
\be
a_1 v_S v_{1x} - 2 a_2 v_{1x}^2 - 2 a_3 v_{2x}^2 = 0.
\label{clmagn-}
\ee
That solution would \emph{not}\/ lead
to the charged-lepton mass matrix being diagonalised by $U_\omega$.

An alternative possibility consists in adding to the theory,
either instead of or in addition to $T^0$,
an alignment field $D^0$ which is a doublet of $S_4$,
has $R$-charge $+2$,
and has $c \left( D^0 \right) = - 2 c \left( S \right)$.
We then have an interaction
\begin{equation}
P_D = D^0 \left( a_5 T_1 T_1 + a_6 T_2 T_2 + a_7 T_1 T_2 \right).
\label{eq:T1T2}
\end{equation}
The ensuing minimisation equations are
\begin{subequations}
\label{eq:D0align}
\begin{eqnarray}
0 = \frac{\partial P_D}{\partial D^0_2} &=&
a_5 \left( v_{1x}^2 + \omega^2 v_{1y}^2 + \omega v_{1z}^2 \right)
+ a_6 \left( v_{2x}^2 + \omega^2 v_{2y}^2 + \omega v_{2z}^2 \right)
\nonumber \\ & &
+ a_7 \left( v_{1x} v_{2x} + \omega^2 v_{1y} v_{2y}
+ \omega v_{1z} v_{2z} \right),
\\
0 = \frac{\partial P_D}{\partial D^0_1} &=&
a_5 \left( v_{1x}^2 + \omega v_{1y}^2 + \omega^2 v_{1z}^2 \right)
+ a_6 \left( v_{2x}^2 + \omega v_{2y}^2 + \omega^2 v_{2z}^2 \right)
\nonumber \\ & &
- a_7 \left( v_{1x} v_{2x} + \omega v_{1y} v_{2y}
+ \omega^2 v_{1z} v_{2z} \right).
\end{eqnarray}
\end{subequations}
These equations are identically satisfied
by the desired alignment in eq.~(\ref{clalign}),
but they display the same type of ambiguity
relative to the sign of the components
already mentioned for the $T^0$ alignment.

If the two alignment methods
(with $T^0$ and $D^0$)
are used together,
then,
for arbitrary values for the parameters $a_{1\mbox{--}7}$
one can only have the alignment
of the desired type---up to the sign ambiguity above
and to related sign ambiguities like $\left( 1, 1, -1 \right)$
or $\left( -1, 1, 1 \right)$.
Indeed,
once we specify $a_{1\mbox{--}4}$ and solve eqs.~(\ref{eq:FT0})
by fixing the relative magnitudes $v_1$ and $v_2$ through eq.~(\ref{clmagn}),
it would take extreme finetuning
for a solution of eqs.~(\ref{eq:D0align}),
which depends on $a_{5\mbox{--}7}$,
to be consistent with that specific solution of eqs.~(\ref{eq:FT0}),
so in general we will be left
only with the solutions that do not depend on $a_{5\mbox{--}7}$,
\textit{viz.}\ with eq.~(\ref{clalign}).
But in fact one of the alignment terms is enough.
Taking \textit{e.g.}\ eqs.~(\ref{eq:D0align})
and expanding to
$v_1 \left( 1, 1 + \mathrm{d}y, 1 + \mathrm{d}z \right)$ for $T_1$
and $v_2 \left( 1, 1 + \mathrm{d}y', 1 + \mathrm{d}z' \right)$ for $T_2$
with infinitesimal perturbations,
one can verify that there are no remaining continuous flat directions
around the minimum.

Noting that the alignment solutions always have two insertions of the flavons,
a simple specific realisation of the auxiliary symmetry
is a $\mathbbm{Z}_3$
with $c \left( \ell_R \right) = 2$ and $c \left( T_1 \right) = 1$.
In this case,
subleading terms could only appear
in eq.~(\ref{eq:PY})
and in the alignment terms with three additional flavon insertions.

\subsection{Neutrino sector}

We want the VEV of $\bar T$
to be aligned in the $\left( k, 1, -1 \right)$ direction,
where $k=-d/f$.\footnote{A bound on $|k|$ may be obtained as follows.
From eqs.~(\ref{abd}),
\be
0 = \sqrt{2} c s e^{- i \beta} k \left( \mu_2 - \mu_3 \right)
+ \left( \mu_1 - s^2 \mu_2 - c^2 \mu_3 \right).
\ee
Therefore,
\be
\label{nnk}
\frac{\left| \left| s^2 m_2 - c^2 m_3 \right| - m_1 \right|}
{\sqrt{2} c s \left( m_2 + m_3 \right)}
\le |k| \le
\frac{s^2 m_2 + c^2 m_3 + m_1}{\sqrt{2} c s \left| m_2 - m_3 \right|}.
\ee
This bound on $|k|$
depends both on $s = \sqrt{3} s_{13}$
and on the neutrino masses.}
According to appendix~\ref{sudzero},
a model with $d=0$ necessitates large neutrino masses
which may or may not conflict with the cosmological bound.
So it is not clear at present whether $k=0$ is possible or should be avoided.
Anyway, we want to have a $\mathbf{3}_1$
aligned in the $\left( k, 1, -1 \right)$ direction
in order to play the role of $\bar T$ in eq.~(\ref{eq:Pnu}).

In order to do this we first introduce the driving fields
in table~\ref{tabled1},
\begin{table}
\centering
\begin{tabular}{|c|cc|cc|}
\hline
Field & $\chi$ & $\theta$ & ${D^0}^\prime$ & ${T^0}^\prime$
\\ \hline
$S_4$ & $\mathbf{3}_2$ & $\mathbf{3}_1$ & $\mathbf{2}$ & $\mathbf{3}_1$
\\
$U(1)_\nu$ & $q \left( \chi \right)$ & $q \left( \theta \right)$ &
$-2 q \left( \chi \right)$ & $-2 q \left( \theta \right)$ \\
$U(1)_R$ & $0$ & $0$ & $2$ & $2$
\\ \hline
\end{tabular} 
\caption{Initial driving fields of the solution for the neutrino sector. \label{tabled1}}
\end{table}
where $q \left( \chi \right)$ and $q \left( \theta \right)$
are generic $U(1)_\nu$ charges.
Then there is a term ${D^0}^\prime \chi \chi$ in the superpotential,
and this term leads,
through equations similar to eqs.~(\ref{eq:D0align}),
to $\left\langle 0 \left| \chi \right| 0 \right\rangle
= v_\chi \left( 1, 1, 1 \right)$.
There is also a term ${T^0}^\prime \theta \theta$ in the superpotential.
If $\left\langle 0 \left| \theta \right| 0 \right\rangle
= \left( v_{\theta x}, v_{\theta y}, v_{\theta z} \right)$,
that term leads to
$v_{\theta y} v_{\theta z} = v_{\theta z} v_{\theta x} = v_{\theta x} v_{\theta y}
= 0$;
we choose $\left\langle 0 \left| \theta \right| 0 \right\rangle$
to lie in the direction $v_\theta \left( 1, 0, 0 \right)$.

We next introduce further driving fields as specified
in table~\ref{drive2}.
\begin{table}
\centering
\begin{tabular}{|c|cc|cc|}
\hline
Field & $s_1$ & $\vartheta$ & $S_1^0$ & $S_2^0$
\\ \hline
$S_4$ & $\mathbf{1}_1$ & $\mathbf{3}_1$ & $\mathbf{1}_1$ & $\mathbf{1}_2$
\\
$U(1)_\nu$ & $q \left( \vartheta \right)$ & $q \left( \vartheta \right)$ &
$-q \left( \vartheta \right) - q \left( \theta \right)$ &
$-q \left( \vartheta \right) - q \left( \chi \right)$
\\
$U(1)_R$ & $0$ & $0$ & $2$ & $2$
\\ \hline
\end{tabular} 
\caption{Further driving fields of the solution for the neutrino sector. \label{drive2}}
\end{table}
These fields allow for terms $S_1^0 \vartheta \theta$ and $S_2^0 \vartheta \chi$
which force the VEV of the triplet $\vartheta$
to be orthogonal to the VEVs of both $\theta$ and $\chi$.
In this way we obtain 
$\left\langle 0 \left| \vartheta \right| 0 \right\rangle
= v_\vartheta \left( 0, 1, -1 \right)$.

Finally,
we assume $- 2 q \left( \vartheta \right)$
to be equal to the $U(1)_\nu$ charge of $D_L \Delta D_L$.
Then,
\be
\bar S =
b_1 s_1 s_1 + b_2 
\left( \vartheta \vartheta \right)_{\mathbf{1}_1},
\quad
\bar D = \left( \vartheta \vartheta \right)_{\mathbf{2}},
\quad
\bar T =
b_3 s_1 \vartheta + b_4
\left( \vartheta \vartheta \right)_{\mathbf{3}_1}
\label{bart2}
\ee
($b_{1\mbox{--}4}$ are coefficients)
are precisely the flavons that we need in eq.~(\ref{eq:Pnu}),
appearing at the same order in flavon insertions.
Indeed,
with $\vartheta = v_\vartheta \left( 0, 1, -1 \right)$,
we have $\left( \vartheta \vartheta \right)_{\mathbf{2}}
= v_\vartheta^2 \left( -1, -1 \right)$
and $\left( \vartheta \vartheta \right)_{\mathbf{3}_1}
= - 2 v_\vartheta^2 \left( 1, 0, 0 \right)$,
according to eq.~(\ref{3times3}).

A simple example for an auxiliary symmetry
realising the assignments listed in the tables
is a $\mathbbm{Z}_8$ with $q \left( D_L \right) = 0$,
$q \left( \Delta \right) = 6$,
and $q \left( \vartheta \right) = 1$.
We may also choose $q \left( \theta \right) = 3$
and $q \left( \chi \right) = 4$.
Then $q \left( S_1^0 \right) = 4$,
$q \left( S_2^0 \right) = 3$,
$q \left( {D^0}^\prime \right) = 0$,
and $q \left( {T^0}^\prime \right) = 2$.
If we assume that
there are no messengers enabling non-renormalisable
alignment terms---see ref.~\cite{Varzielas:2012ai}---then
the concern is with the mass terms,
and in this specific case all the subleading contributions
to eq.~(\ref{bart2}) appear with at least two additional fields---for instance,
one may add $\left( \chi \chi \right)_{\mathbf{1}_1}$
to any of the combinations in eq.~(\ref{bart2}).
Those subleading contributions have
a suppression by at least two extra powers of $\Lambda$ and,
provided $\Lambda$ is much larger than the VEVs of the flavons,
they are expected to be negligible.

On the other hand,
if non-renormalisable alignment terms are allowed,
then with this $\mathbbm{Z}_8$ there are problematic alignment terms
which appear with one additional field insertion.
In order to disallow them one may use
a symmetry $\mathbbm{Z}_{14}$ instead of $\mathbbm{Z}_8$;
there are then several possibilities for the charges $q \left( \theta \right)$
and $q \left( \chi \right)$---one possibility is $q \left( \Delta \right) = 12$,
$q \left( \vartheta \right) = 1$,
$q \left( \theta \right) = 3$,
and $q \left( \chi \right) = 7$,
and then $q \left( S_1^0 \right) = 10$,
$q \left( S_2^0 \right) = 6$,
$q \left( {D^0}^\prime \right) = 0$,
and $q \left( {T^0}^\prime \right) = 8$.

\section{Conclusions \label{sec:conc}}

In this paper we have considered
the phenomenological consequences of the TM$_1$ \textit{Ansatz}
in light of the recent experimental data
on a nonzero reactor mixing angle
and on a non-maximal atmospheric mixing angle.
We have provided an explicit framework,
based on a lepton flavour symmetry group $S_4$,
for models with TM$_1$ mixing.
Confronting the predictions of some of those models with the experimental data
may rule out or constrain these particular cases.
We have investigated how the VEVs of the required $S_4$ multiplets
can be aligned consistently.

\vspace*{3mm}

\paragraph{Acknowledgements:}
IdMV would like to thank CFTP at Instituto Superior T\'ecnico
for hospitality during part of this work.
LL thanks A.~Yu.~Smir\-nov and D.~Hern\'andez for having organised
the very enjoyable BENE 2012 workshop at ICTP,
at which a preliminary version of this work has been presented.
The work of IdMV was supported by the grant PA 803/6-1
of the \textit{Deutsche Forschungsgemeinschaft}.
The work of LL is supported by Portuguese national funds through
\textit{Funda\c c\~ao para a Ci\^encia e a Tecnologia}\/ (FCT)
project PEst-OE/FIS/UI0777/2011,
and also through
the Marie Curie Initial Training Network ``UNILHC'' PITN-GA-2009-237920
and
the project CERN/FP/123580/2011.
Both authors are partially supported by FCT
through the project PTDC/FIS/098188/2008.

\vspace*{5mm}

\begin{appendix}

\setcounter{equation}{0}
\renewcommand{\theequation}{A\arabic{equation}}

\section{The group $S_4$}

The group $S_4$ is the group of permutations of four objects $o_{1,2,3,4}$.
It is generated by two permutations,
$f: o_1 \leftrightarrow o_2$ and $g: o_2 \to o_3 \to o_4 \to o_2$.
Those permutations satisfy
\begin{equation}
f^2 = g^3 = \left( f g \right)^4 = e,
\end{equation}
where $e$ is the identity permutation.
The group $S_4$ has order $4! = 24$ and five
inequivalent irreducible representations (`irreps'):
the triplets $\mathbf{3}_1$ and $\mathbf{3}_2$,
the doublet $\mathbf{2}$,
and the singlets $\mathbf{1}_1$ and $\mathbf{1}_2$.
The $\mathbf{1}_1$ is the trivial representation.
The $\mathbf{1}_2$ makes $f \to -1$,
$g \to +1$.
We choose a basis for the doublet such that
\begin{equation}
\mathbf{2}: \quad
f \to F_2 = \left( \begin{array}{cc} 0 & 1 \\ 1 & 0 \end{array} \right), \quad
g \to G_2 = \left( \begin{array}{cc} \omega & 0 \\ 0 & \omega^2
\end{array} \right),
\end{equation}
where $\omega = \exp{\left( i 2 \pi / 3 \right)}$.
Notice that this representation is not faithful,
since $\left( F_2 G_2 \right)^2$ already is the unit matrix.\footnote{The
$\mathbf{2}$ is a faithful representation of the $S_3$ subgroup of $S_4$
formed by the permutations of $o_{2,3,4}$.}
For the $\mathbf{3}_1$ we choose a basis such that
\begin{equation}
\label{G3F3}
\mathbf{3}_1: \quad
f \to F_3 = \left( \begin{array}{ccc} 1 & 0 & 0 \\ 0 & 0 & -1 \\ 0 & -1 & 0
\end{array} \right), \quad
g \to G_3 = \left( \begin{array}{ccc} 0 & 1 & 0 \\ 0 & 0 & 1 \\ 1 & 0 & 0
\end{array} \right).
\end{equation}
Notice that $\mathbf{1}_2 = \det{\mathbf{3}_1}$.
The $\mathbf{3}_2 = \mathbf{3}_1 \otimes \mathbf{1}_2$,
therefore
\begin{equation}
\mathbf{3}_2: \quad f \to - F_3, \quad g \to G_3.
\end{equation}
The irreps $\mathbf{3}_1$ and $\mathbf{3}_2$ are faithful.
Notice that the matrices of the $\mathbf{3}_2$ have determinant $+1$,
therefore $S_4$ is isomorphic to a subgroup of $SO(3)$.
That subgroup is the symmetry group
of the cube or of the regular octahedron.

Let $\left( x, y, z \right)$ and $\left( x', y', z'\right)$
be two identical triplets of $S_4$,
either both of them $\mathbf{3}_1$ or both of them $\mathbf{3}_2$.
Then,
\begin{eqnarray}
& &
\left( \begin{array}{c}
y z' + z y' \\ z x' + x z' \\ x y' + y x'
\end{array} \right),
\
\left( \begin{array}{c}
y z' - z y' \\ z x' - x z' \\ x y' - y x'
\end{array} \right),
\nonumber \\ & &
\left( \begin{array}{c}
x x' + \omega^2 y y' + \omega z z' \\ x x' + \omega y y' + \omega^2 z z'
\end{array} \right),
\
x x'+ y y' + z z'
\label{3times3}
\end{eqnarray}
are,
respectively,
a $\mathbf{3}_1$,
a $\mathbf{3}_2$,
a $\mathbf{2}$,
and a $\mathbf{1}_1$.
If,
on the other hand,
$\left( x, y, z \right)$ is a $\mathbf{3}_1$
and $\left( x', y', z'\right)$ a $\mathbf{3}_2$ of $S_4$,
then
\begin{eqnarray}
& &
\left( \begin{array}{c}
y z' - z y' \\ z x' - x z' \\ x y' - y x'
\end{array} \right),
\
\left( \begin{array}{c}
y z' + z y' \\ z x' + x z' \\ x y' + y x'
\end{array} \right),
\nonumber \\ & &
\left( \begin{array}{c}
x x' + \omega^2 y y' + \omega z z' \\ - x x' - \omega y y' - \omega^2 z z'
\end{array} \right),
\
x x'+ y y' + z z'
\end{eqnarray}
are a $\mathbf{3}_1$,
a $\mathbf{3}_2$,
a $\mathbf{2}$,
and a $\mathbf{1}_2$,
respectively.

Let $\left( x, y, z \right)$ be a $\mathbf{3}_1$
and $\left( p, q \right)$ a $\mathbf{2}$ of $S_4$.
Then,
\begin{equation}
\left( \begin{array}{c}
x \left( p + q \right) \\
y \left( \omega p + \omega^2 q \right) \\
z \left( \omega^2 p + \omega q \right)
\end{array} \right)
\quad \mbox{and} \quad
\left( \begin{array}{c}
x \left( p - q \right) \\
y \left( \omega p - \omega^2 q \right) \\
z \left( \omega^2 p - \omega q \right)
\end{array} \right)
\label{mult2}
\end{equation}
are a $\mathbf{3}_1$ and a $\mathbf{3}_2$,
respectively.
If,
however,
the $\left( x, y, z \right)$ were a $\mathbf{3}_2$,
then the multiplets in eq.~(\ref{mult2})
would be a $\mathbf{3}_2$ and a $\mathbf{3}_1$,
respectively.

Let $\left(p, q \right)$ and $\left( p', q' \right)$ be two $\mathbf{2}$
of $S_4$,
then
\begin{equation}
\left( \begin{array}{c} q q' \\ p p' \end{array} \right),
\quad
p q'+ q p', \quad p q'- q p'
\end{equation}
are a $\mathbf{2}$,
a $\mathbf{1}_1$,
and a $\mathbf{1}_2$,
respectively.

If $t$ transforms as a $\mathbf{1}_2$ of $S_4$
and $\left( x, y, z \right)$ is either a $\mathbf{3}_1$ or a $\mathbf{3}_2$,
then $\left( t x, t y, t z \right)$ will correspondingly be
either a $\mathbf{3}_2$ or a $\mathbf{3}_1$,
respectively.
If $\left( p, q \right)$ is a $\mathbf{2}$,
then $\left( t p, - t q \right)$ is also a $\mathbf{2}$.

\setcounter{equation}{0}
\renewcommand{\theequation}{B\arabic{equation}}

\section{Possibilities with one vanishing parameter}
\label{sub1zero}

In this appendix we investigate the cases where either $a$,
$b$,
or $d$ vanish.
In practice we shall have to deal with equations of the form
\be
\label{sum}
\mu_1 + p \mu_2 + q \mu_3 = 0,
\ee
where $p$ and $q$ are complex numbers with known moduli,
$|p|^2 \equiv P$ and $|q|^2 \equiv Q$.
Unfortunately,
eq.~(\ref{sum}) is not very well defined
because the $\mu_j = m_j e^{- i 2 \psi_j}$
contain unknown Majorana phases $\psi_j$;
moreover,
the phases of $p$ and $q$ are in some cases also ambiguous.
Equation~(\ref{sum}) states that the three complex numbers $\mu_1$,
$p \mu_2$,
and $q \mu_3$ form a triangle in the complex plane.
We make use of the fact~\cite{book} that,
if three complex numbers $c_{1,2,3}$ form a triangle in the complex plane,
then the moduli of those numbers must satisfy the inequality
\be
\label{eq:lambda}
\left| c_1 \right|^4 + \left| c_2 \right|^4 + \left| c_3 \right|^4
- 2 \left| c_1 c_2 \right|^2 - 2 \left| c_1 c_3 \right|^2
- 2 \left| c_2 c_3 \right|^2 \le 0.
\ee
This allows us to extract a useful inequality from eq.~(\ref{sum}):
\be
\label{ineq}
\lambda \equiv
m_1^4 + P^2 m_2^4 + Q^2 m_3^4 - 2 P m_1^2 m_2^2 - 2 Q m_1^2 m_3^2
- 2 P Q m_2^2 m_3^2 \le 0.\
\ee
As we shall see,
this inequality allows us to dismiss some of the models
and to constrain the neutrino masses in the remaining ones.

If the neutrino mass hierarchy is normal,
then
\be
\label{normal}
m_2^2 = m_1^2 + \Delta m^2_\mathrm{sol}, \quad
m_3^2 = m_1^2 + \Delta m^2_\mathrm{atm},
\ee
where $\Delta m^2_\mathrm{sol} \approx 7.5 \times 10^{-5}\, \mathrm{eV}^2$
and $\Delta m^2_\mathrm{atm} \approx 2.4 \times 10^{-3}\, \mathrm{eV}^2$ are,
respectively,
the solar and atmospheric neutrino mass-squared differences.
On the other hand,
if the neutrino mass hierarchy is inverted,
then
\be
\label{inverted}
m_1^2 = m_3^2 + \Delta m^2_\mathrm{atm}, \quad
m_2^2 = m_3^2 + \Delta m^2_\mathrm{atm} + \Delta m^2_\mathrm{sol}.
\ee
We know that
\be
\label{epsilon}
\epsilon \equiv \frac{\Delta m^2_\mathrm{sol}}{\Delta m^2_\mathrm{atm}}
\ee
is very small,
$0.0294 \le \epsilon \le 0.0335$~\cite{fogli}.
Another small quantity,
of the same order of magnitude as $\epsilon$,
is $s^2$,
which,
as follows from eq.~(\ref{s}),
satisfies $0.0648 \le s^2 \le 0.0798$ at the $1 \sigma$ level.
A third small quantity is
\be
X \equiv 4 c^2 s^2 \sin^2{\beta}
= \frac{3}{2}\, c_{13}^4 \left( 1 - 2 s_{23}^2 \right)^2,
\ee
which phenomenologically may be as large as 0.18
but may conceivably be much smaller than that,
or even zero~\cite{forero,fogli,schwetz}.

In the case of a normal neutrino mass spectrum,
$m_1$ is the smallest neutrino mass;
let then $f_1 \equiv m_1^2 / \Delta m^2_\mathrm{atm}$.
In that case,
from eqs.~(\ref{ineq}),
(\ref{normal}),
and~(\ref{epsilon}),
\ba
\frac{\lambda_\mathrm{normal}}{\left( \Delta m^2_\mathrm{atm} \right)^2} &=&
f_1^2 \left( 1 + P^2 + Q^2 - 2 P - 2 Q - 2 P Q \right)
\no & &
+ 2 f_1 \left[ Q \left( Q - 1 - P \right)
+ \epsilon P \left( P - 1 - Q \right) \right]
\no & &
+ \left( Q - \epsilon P \right)^2 \le 0.
\label{f1}
\ea
In the case of an inverted neutrino mass spectrum,
$m_3$ is the smallest neutrino mass;
let in that case $f_3 \equiv m_3^2 / \Delta m^2_\mathrm{atm}$.
Then,
from eqs.~(\ref{ineq}),
(\ref{inverted}),
and~(\ref{epsilon}),
\ba
\frac{\lambda_\mathrm{inverted}}{\left( \Delta m^2_\mathrm{atm} \right)^2} &=&
f_3^2 \left( 1 + P^2 + Q^2 - 2 P - 2 Q - 2 P Q \right)
\no & &
+ 2 f_3 \left[ 1 - P - Q
+ \left( 1 + \epsilon \right) P \left( P - 1 - Q \right) \right]
\no & &
+ \left( 1 - P - \epsilon P \right)^2 \le 0.
\label{f3}
\ea

\subsection{The case $d=0$ \label{sudzero}}

\begin{figure}
\begin{center}
\epsfig{file=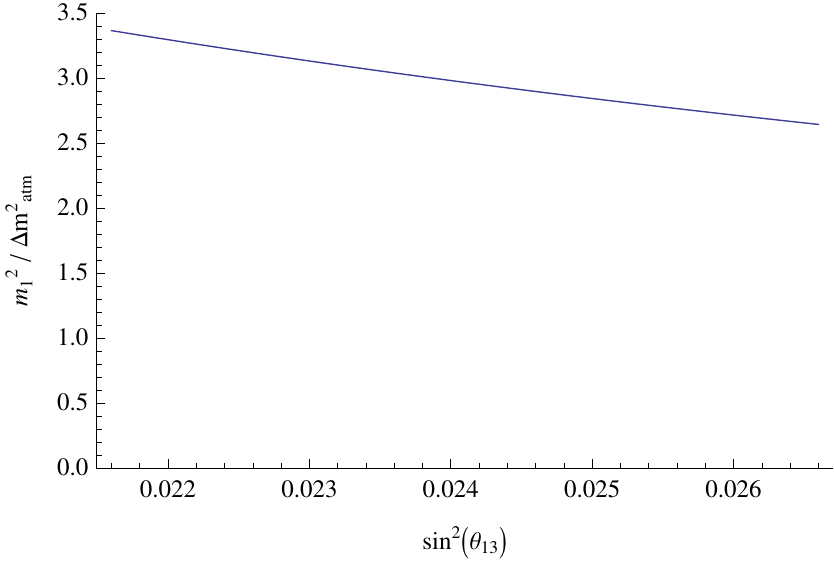,width=0.75\textwidth}
\end{center}
\caption{The minimum value for $m_1^2 / \Delta m^2_\mathrm{atm}$
as a function of $s_{13}^2$
when $d = 0$.}
We have used the best-fit value of
$\Delta m^2_\mathrm{sol} / \Delta m^2_\mathrm{atm}$
in drawing this figure.
\label{fig:dzeroNH}
\end{figure}

We start with the simplest case,
namely $d=0$.
Then,
from eqs.~(\ref{abd}),
\be
\label{rel1}
\mu_1 = s^2 \mu_2 + c^2 \mu_3.
\ee
This means that $P = s^4$ and $Q = c^4$.
Then,
$1 + P^2 + Q^2 - 2 P - 2 Q - 2 P Q = 0$.
One quickly finds that \emph{an inverted neutrino mass spectrum
is not possible} in this case,
while a normal neutrino mass spectrum is possible provided
\be
f_1 \ge \frac{\left( c^4 - \epsilon s^4 \right)^2}
{4 c^2 s^2 \left( c^2 + \epsilon s^2 \right)}
\approx \frac{1}{12 s_{13}^2}.
\ee
This inequality corresponds approximately
to the area above the line in fig.~\ref{fig:dzeroNH}
(where the best-fit value for $\epsilon$ was used).

Therefore,
if the neutrino mass hierarchy is normal,
\be
m_1^2 \ge
\frac{\left( c^4 \Delta m^2_{\mathrm{atm}}
- s^4 \Delta m^2_{\mathrm{sol}} \right)^2}{4 c^2 s^2
\left( c^2 \Delta m^2_{\mathrm{atm}}
+ s^2 \Delta m^2_{\mathrm{sol}} \right)}
\approx \frac{c^4}{4 s^2}\, \Delta m^2_{\mathrm{atm}}
\approx 0.00718\, \mathrm{eV}^2,
\ee
or $m_1 \ge 0.085\, \mathrm{eV}$;
this indicates an almost-degenerate neutrino mass spectrum.
Then the sum of the light-neutrino masses is
\be
m_1 + m_2 + m_3 \ge
\left( \sqrt{0.00718} + \sqrt{0.00720} + \sqrt{0.00959} \right) \mathrm{eV}
\approx 0.268\, \mathrm{eV}.
\ee
This is a value which violates a recent cosmological bound~\cite{nubound}
but fits nicely with a different one~\cite{cosmology2}.

\subsection{The case $a=0$ \label{suazero}}

\begin{figure}
\begin{center}
\epsfig{file=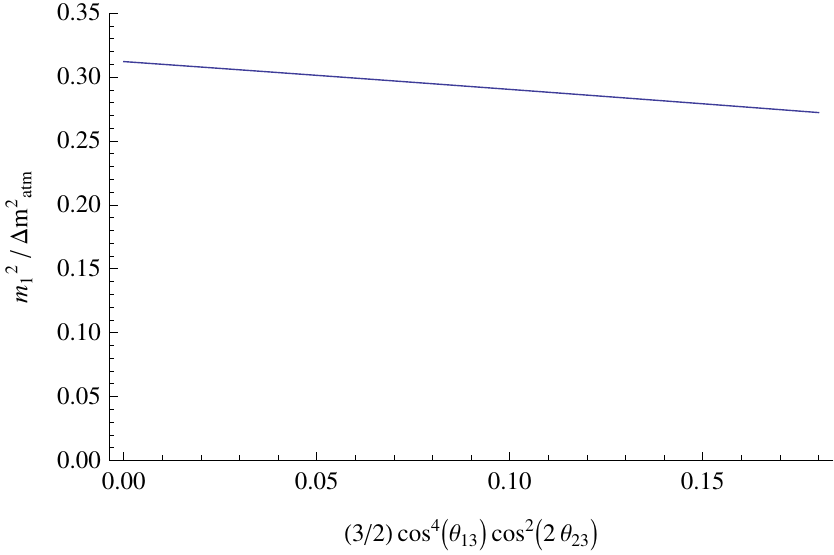,width=0.75\textwidth}
\end{center}
\caption{The minimum value for $m_1^2 / \Delta m^2_\mathrm{atm}$ in the case $a=0$
as a function of $X=(3/2)\, \cos^4(\theta_{13}) \cos^2(2 \theta_{23})$,
for the best-fit values of $\epsilon$ and $s_{13}^2$.}
\label{fig:azeroNH}
\end{figure}

\begin{figure}
\begin{center}
\epsfig{file=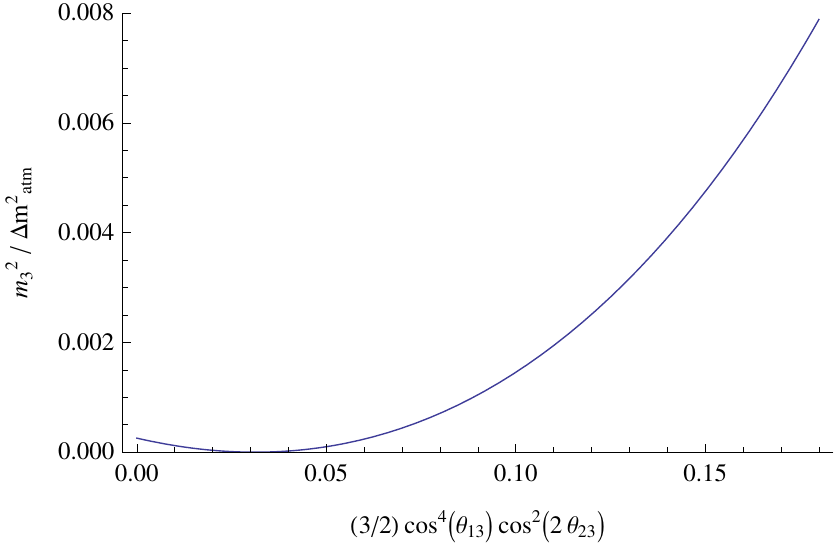,width=0.75\textwidth}
\end{center}
\caption{The minimum value for $m_3^2 / \Delta m^2_\mathrm{atm}$ in the case $a=0$
as a function of $X=(3/2)\, \cos^4(\theta_{13}) \cos^2(2 \theta_{23})$,
for the best-fit values of $\epsilon$ and $s_{13}^2$.}
\label{fig:azeroIH}
\end{figure}

We now consider the case $a=0$.
It then follows from eqs.~(\ref{abd}) that
\be
\label{real2}
\mu_1 + \left( s^2 + c^2 e^{- i 2 \beta} \right) \mu_2
+ \left( c^2 + s^2 e^{- i 2 \beta} \right) \mu_3 = 0.
\ee
Hence,
in this case
\be
P = Q = 1 - X
\ee
and
\begin{subequations}
\ba
\frac{- \lambda_\mathrm{normal}}{\left( \Delta m^2_\mathrm{atm} \right)^2} &=&
f_1^2 \left( 3 - 4 X \right)
+ 2 f_1 P \left( 1 + \epsilon \right)
\no & &
- P^2 \left( 1 - \epsilon \right)^2 \ge 0,
\\
\frac{- \lambda_\mathrm{inverted}}{\left( \Delta m^2_\mathrm{atm} \right)^2} &=&
f_3^2 \left( 3 - 4 X \right)
+ 2 f_3 \left( 2 - 3 X + \epsilon - \epsilon X \right)
\no & &
- \left( X - \epsilon + \epsilon X \right)^2 \ge 0.
\ea
\end{subequations}
These inequalities yield
\begin{subequations}
\ba
f_1 &\ge& \frac{1 - X}{3 - 4 X} \left[
2 \sqrt{ \left( 1 + \epsilon^2 \right) \left( 1 - X \right)
- \epsilon \left( 1 - 2 X \right)} - 1 - \epsilon \right],
\hspace*{7mm}
\\
f_3 &\ge& \frac{1 - X}{3 - 4 X} \left[
2 \sqrt{ \left( 1 + \epsilon^2 \right) \left( 1 - X \right)
+ \epsilon \left( 1 - 2 X \right)} - 2 - \epsilon
\right. \no & & \left.
+ \frac{X}{1 - X} \right],
\ea
\end{subequations}
which are valid for the normal- and inverted-hierarchy cases,
respectively.
So,
in the case of a normal hierarchy
$m_1^2 \gtrsim  \Delta m^2_\mathrm{atm} / 3$
while in the case of an inverted hierarchy
$m_3^2 \gtrsim  \left( X - \epsilon \right)^2 \Delta m^2_\mathrm{atm} / 4$
may be considerably smaller ($m_3$ may even vanish if it happens that
$X - \epsilon + \epsilon X = 0$).
This can be seen in figs.~\ref{fig:azeroNH} and~\ref{fig:azeroIH},
respectively.

In this case both neutrino mass spectra are allowed.

\subsection{The case $b=0$ \label{subzero}}

\begin{figure}
\begin{center}
\epsfig{file=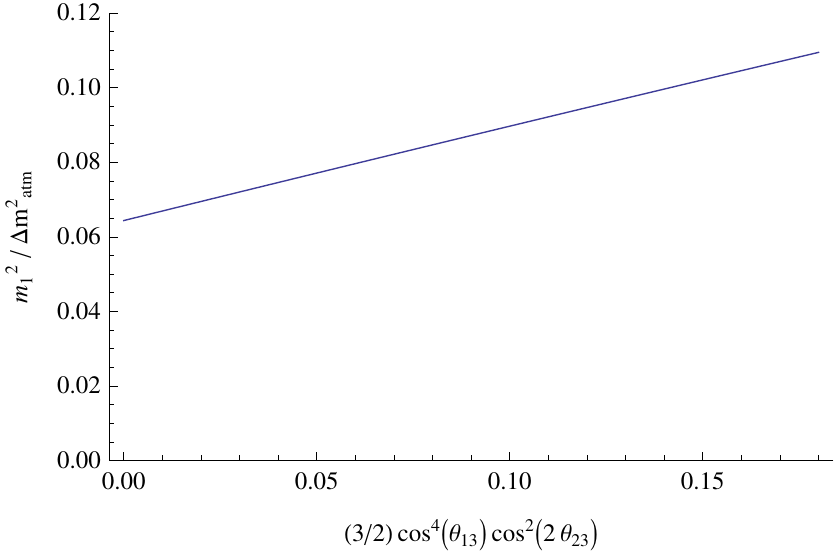,width=0.75\textwidth}
\end{center}
\caption{The minimum value for $m_1^2 / \Delta m^2_\mathrm{atm}$ in the case $b=0$
as a function of $X=(3/2)\, \cos^4(\theta_{13}) \cos^2(2 \theta_{23})$,
for the best-fit values of $\epsilon$ and $s_{13}^2$.}
\label{fig:bzeroNH}
\end{figure}

\begin{figure}
\begin{center}
\epsfig{file=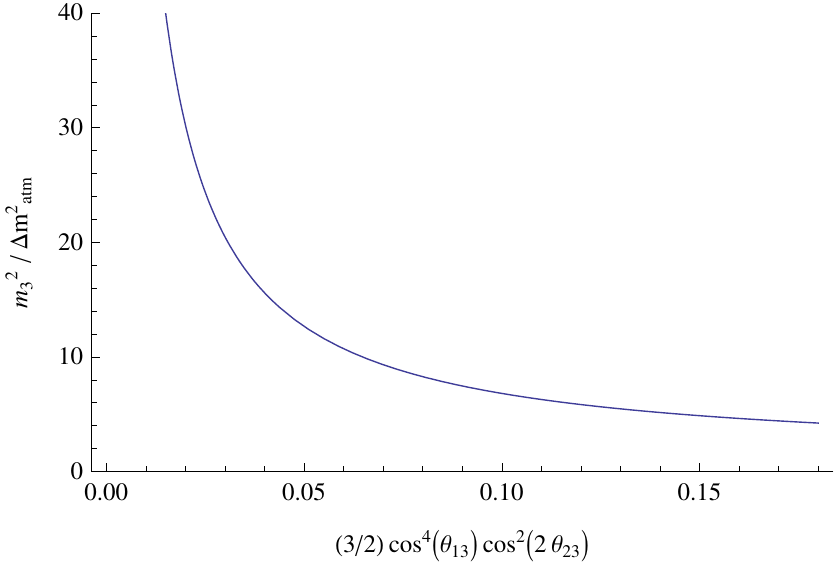,width=0.75\textwidth}
\end{center}
\caption{The minimum value for $m_3^2 / \Delta m^2_\mathrm{atm}$ in the case $b=0$
as a function of $X=(3/2)\, \cos^4(\theta_{13}) \cos^2(2 \theta_{23})$,
for the best-fit values of $\epsilon$ and $s_{13}^2$.}
\label{fig:bzeroIH}
\end{figure}

Equations~(\ref{abd}) in this case lead to
\be
\label{real3}
\mu_1 + \left( s^2 - 2 c^2 e^{- i 2 \beta} \right) \mu_2
+ \left( c^2 - 2 s^2 e^{- i 2 \beta} \right) \mu_3 = 0.
\ee
Therefore,
\begin{subequations}
\ba
P &=& 4 - 12 s^2 + 9 s^4 + 2 X
\no &=& \left( 2 - 3 s^2 \right)^2 + 2 X,
\\
Q &=& 1 - 6 s^2 + 9 s^4 + 2 X
\no &=& \left( 1 - 3 s^2 \right)^2 + 2 X.
\ea
\end{subequations}
The inequalities~(\ref{f1}) and~(\ref{f3}) then read
\begin{subequations}
\ba
8 X f_1^2 & &
\no
+ 4 f_1 \left\{
\left( 1 - 3 s^2 \right) \left( 2 - 3 s^2 \right)
\left[ 1 - 3 s^2 - \epsilon \left( 2 - 3 s^2 \right) \right]
\right. & &
\no
\left. + 2 X \left[ 2 - 3 s^2 - \epsilon \left( 1 - 3 s^2 \right) \right] 
\right\} & &
\no
- \left[ \left( 1 - 3 s^2 \right)^2 + 2 X - \epsilon \left( 2 - 3 s^2 \right)^2
- 2 \epsilon X \right]^2
&\ge& 0, \hspace*{5mm}
\\
8 X f_3^2 & &
\no
- 4 f_3 \left\{
\left( 1 - 3 s^2 \right) \left( 2 - 3 s^2 \right) \left[
\left( 1 - 3 s^2 \right) + \epsilon \left( 2 - 3 s^2 \right) \right]
\right. & & \no
\left. + 2 X \left( \epsilon - 3 s^2 - 3 \epsilon s^2 \right)
\right\} & &
\no
- \left[ - 3 - 4 \epsilon + \left( 1 + \epsilon \right)
\left( 12 s^2 - 9 s^4 - 2 X \right) \right]^2
&\ge& 0,
\label{f32}
\ea
\end{subequations}
respectively.
The corresponding lower bounds on $f_1$ and $f_3$ are depicted
in figs.~\ref{fig:bzeroNH} and~\ref{fig:bzeroIH},
respectively.
One sees that a normal mass spectrum is allowed
but that,
unless $X \gtrsim 0.05$,
a inverted mass spectrum leads to much too high neutrino masses,
which violate the cosmological bound.

\end{appendix}

\end{document}